\documentclass[conference]{IEEEtran}
\IEEEoverridecommandlockouts
\usepackage{cite}
\usepackage{amsmath,amssymb,amsfonts}
\usepackage{algorithmic}
\usepackage{graphicx}
\usepackage{listings}
\usepackage{textcomp}
\usepackage{xcolor}
\usepackage{hyperref}
\usepackage{booktabs}
\usepackage{paralist}
\usepackage[ancient]{flushend}

\hypersetup{
  colorlinks,
  breaklinks,
  linkcolor={green!80!black},
  citecolor={red!70!black},
  urlcolor={blue!70!black}
}

\lstdefinelanguage[Arm]{Assembler}[x86masm]{Assembler}{%
  morekeywords = {%
    ldr,movw,movt,tbb,tbh,adr.w,add.w,mov,b.w,%
    .byte,.short,.word,%
  },%
}

\def\BibTeX{{\rm B\kern-.05em{\sc i\kern-.025em b}\kern-.08em
    T\kern-.1667em\lower.7ex\hbox{E}\kern-.125emX}}

\def\System{PicoXOM}

\def\XOR{\texorpdfstring{$\oplus$}{\^{}}}

\begin{document}

\title{Fast Execute-Only Memory for Embedded Systems}

\author{
  Zhuojia Shen \\
  {\it Department of Computer Science} \\
  {\it University of Rochester} \\
  Rochester, NY \\
  zshen10@cs.rochester.edu
  \and
  Komail Dharsee \\
  {\it Department of Computer Science} \\
  {\it University of Rochester} \\
  Rochester, NY \\
  kdharsee@cs.rochester.edu
  \and
  John Criswell \\
  {\it Department of Computer Science} \\
  {\it University of Rochester} \\
  Rochester, NY \\
  criswell@cs.rochester.edu
}

\maketitle

\begin{abstract}
Remote code disclosure attacks threaten embedded systems as
they allow attackers to steal intellectual property or
to find reusable code for use in control-flow hijacking attacks.
Execute-only memory (XOM)
prevents remote code disclosures, but existing XOM solutions either
require a memory management unit that is not available
on ARM embedded systems or incur significant overhead.

We present \emph{\System}: a fast and novel XOM system for
ARMv7-M and ARMv8-M devices which leverages ARM's
Data Watchpoint and Tracing unit along with the processor's
simplified memory protection hardware.  On average,
{\System} incurs 0.33\% performance overhead and 5.89\% code size
overhead on two benchmark suites and five real-world applications.
\end{abstract}

\section{Introduction}
\label{sec:intro}

Remote code disclosure attacks threaten computer systems.  Remote attackers
exploiting buffer overread vulnerabilities~\cite{Overread:EuroSec09} can not
only steal intellectual property (e.g., proprietary application code, for reverse
engineering), but also leak code to locate gadgets for
advanced code reuse attacks~\cite{JIT-ROP:Oakland13}, thwarting code layout
diversification defenses like Address Space Layout Randomization
(ASLR)~\cite{ASLR:PaX01}.
Embedded Internet-of-Things (IoT) devices exacerbate the situation; many of these
microcontroller-based systems have the same Internet connectivity as
desktops and servers but rarely employ protections against
attacks~\cite{IoT:DAC15,ThermoSpy:BlackHatUSA14}.
Given the ubiquity of these embedded devices in industrial production
and in our lives, making them immune
to code disclosure attacks is crucial.

Recent research~\cite{XnR:CCS14,HideM:CODASPY15,Readactor:Oakland15,%
LR2:NDSS16,ExOShim:ICCWS16,KHide:CNS16,NORAX:Oakland17,kR^X:EuroSys17,%
XOM-Switch:BlackHatAsia18,IskiOS:ArXiv19,uXOM:UsenixSec19} implements
\emph{execute-only memory (XOM)} to defend against code disclosure
attacks.  Despite being unable to prevent code pointer leakage from
data regions such as heaps and stacks, XOM enforces memory
protection on the code region so that
instruction fetching is allowed but reading or writing instructions as
data is disallowed.  This simple and effective defense, however, is not
natively available on low-end microcontrollers.  For example, the
ARMv7-M and ARMv8-M architectures used in mainstream devices support memory
protection but not
execute-only (XO) permissions~\cite{ARMv7-M:Manual,ARMv8-M:Manual}.
uXOM~\cite{uXOM:UsenixSec19} implements
XOM on ARM embedded systems but incurs
significant performance and code size overhead (7.3\% and 15.7\%,
respectively) as it
transforms most load instructions into special unprivileged load
instructions.  Given embedded systems' real-time constraints and limited
memory resources, a practically ideal XOM solution should have
\emph{close-to-zero performance penalty} and
\emph{minimal memory overhead}.

This paper presents \emph{\System}, a fast and novel XOM
system for ARMv7-M and ARMv8-M devices using a memory protection unit
(MPU) and the Data Watchpoint and Tracing (DWT)
unit~\cite{ARMv7-M:Manual,ARMv8-M:Manual}.  {\System} uses the MPU to enforce
\emph{write} protection on code and uses the
unique \emph{address range matching capability}
of the DWT unit to control \emph{read} access to the code region.  On a
matched access, the DWT unit generates a debug monitor exception
indicating an illegal code read, while unmatched accesses execute
normally without slowdown. As {\System} disallows all read
accesses to the code segment, it includes a minimal compiler change that
removes all data embedded in the code segment.

We built a prototype of {\System} and
evaluated it on an ARMv7-M board with two benchmark suites and five
real-world embedded applications.  Our results show that {\System} adds
\emph{negligible performance overhead} of 0.33\% and only has a
\emph{small code size
increase} of 5.89\% while providing strong protection against code disclosure
attacks.

To summarize, our contributions are:
\begin{itemize}
\item
  {\System}: a novel method of utilizing the ARMv7-M and ARMv8-M
  debugging facilities to implement XOM.
  To the best of our knowledge, this is the first use of ARM debug
  features for security purposes.

\item
  A prototype implementation of {\System} on ARMv7-M.

\item
  An evaluation of {\System}'s performance and code size impact on the
  BEEBS benchmark suite, the CoreMark-Pro benchmark suite, and five
  real-world embedded applications, showing that {\System} only incurs 0.33\%
  run-time overhead and 5.89\% code size overhead.
\end{itemize}

The rest of the paper is organized as follows.  Section~\ref{sec:bg}
provides background information on ARMv7-M and ARMv8-M.
Section~\ref{sec:threat} describes our threat model and assumptions.
Sections~\ref{sec:design} and \ref{sec:impl} present the design and
implementation of {\System}, respectively.  Section~\ref{sec:eval}
reports on our evaluation of {\System}, Section~\ref{sec:related}
discusses related work, and Section~\ref{sec:conc} concludes the paper
and discusses future work.

\section{Background}
\label{sec:bg}

{\System} targets ARMv7-M and ARMv8-M architectures, which cover a wide
range of embedded devices on the market, and it leverages unique
features of these architectures.  This section provides
important background material on the instruction sets, execution modes,
address space layout, memory protection mechanisms, and on-chip debug
support found in ARMv7-M and ARMv8-M.

\subsection{Instruction Sets and Execution Modes}
\label{sec:bg:isa}

ARMv7-M~\cite{ARMv7-M:Manual} and ARMv8-M~\cite{ARMv8-M:Manual} are the
mainstream M-profile ARM architectures for embedded microcontrollers.
Unlike ARM's A and R profiles, they only support the Thumb instruction
set which is a mixture of 16-bit and 32-bit densely-encoded Thumb
instructions.

ARMv7-M~\cite{ARMv7-M:Manual} supports two execution modes:
unprivileged mode and
privileged mode.  An ARMv7-M processor always executes
exception handlers in privileged mode, while application code is allowed
to execute in either mode.  Code running in unprivileged mode can raise the
current execution mode to
privileged mode using a supervisor call instruction ({\tt SVC}).
This is typically how ARMv7-M realizes system calls.  However,
embedded applications usually run in privileged mode to reduce
the cost of system calls.

ARMv8-M inherits all the features of ARMv7-M and adds a security
extension called TrustZone-M~\cite{ARMv8-M:Manual} that isolates
software into a secure world and a non-secure world; this effectively
doubles the execution modes as software can be executing in either world,
privileged or unprivileged.

\subsection{Address Space Layout}
\label{sec:bg:layout}

\begin{figure}[tb]
  \centering
  \resizebox{1.0\columnwidth}{!}{%
    \includegraphics{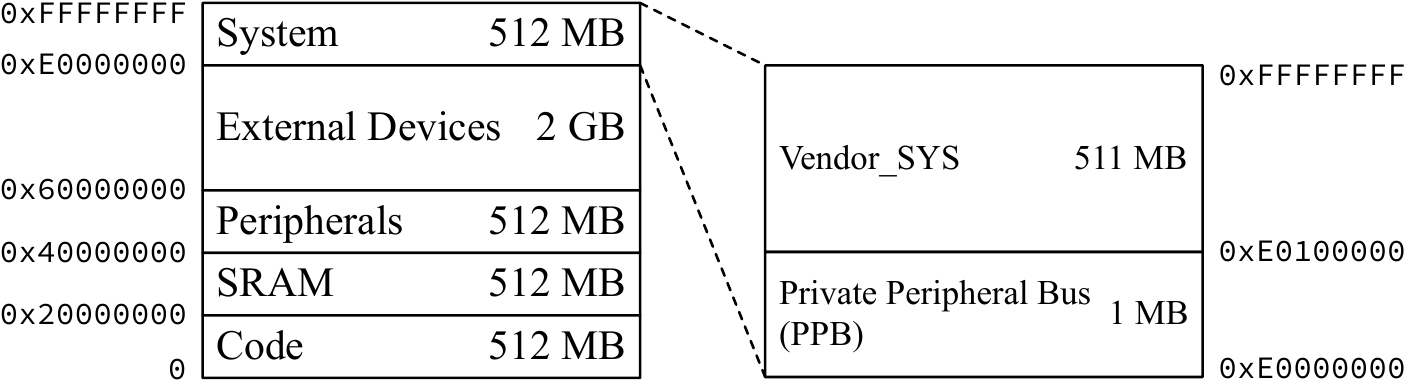}
  }
  \caption{Memory Layout of ARMv7-M and ARMv8-M Architectures}
  \label{fig:address-space}
\end{figure}

Both ARMv7-M~\cite{ARMv7-M:Manual} and ARMv8-M~\cite{ARMv8-M:Manual}
architectures operate on a single 32-bit
physical address space and use memory-mapped I/O to access external devices and
peripherals.  As Figure~\ref{fig:address-space} shows, the address space
is generally divided into eight consecutive 512~MB regions; the
{\tt Code} region maps flash memory/ROM that contains code and read-only
data, the {\tt SRAM} region typically contains heaps and stacks, and
the {\tt System} region holds memory-mapped system registers including a
Private Peripheral Bus (PPB) subregion.  The PPB subregion contains all
critical system registers such as MPU configuration registers and the
Vector Table Offset Register {\tt VTOR}.
All other regions are for
memory-mapped peripherals and external devices.
Note that ARMv7-M and ARMv8-M do not have special privileged
instructions to access system registers mapped in the {\tt System}
region; instead, they can be modified by regular load and store
instructions.

\subsection{Memory Protection Unit}
\label{sec:bg:mpu}

ARMv7-M and ARMv8-M devices do not have a memory management unit (MMU)
that supports virtual memory; instead, they support an
optional MPU that can be configured to enforce
region-based access control on physical
memory~\cite{ARMv7-M:Manual,ARMv8-M:Manual}.  A typical ARMv7-M device
supports up to 8 MPU regions, each of which is configurable with a
base address, a power-of-two size from 32~bytes to 4~GB, and separate access
permissions (R, W, and X) for privileged and unprivileged
modes.  With TrustZone-M, ARMv8-M has separate MPU
configurations for secure and non-secure worlds~\cite{ARMv8-M:Manual}.
MPU configuration registers are in the PPB region.

There are, however, limitations on how one can configure access permissions
for an MPU region.  First, the privileged access permission cannot be
more restrictive than the unprivileged one; this prohibits an MPU region
with, for example, unprivileged read-write and privileged read-only
permissions.  Second, the PPB region is always privileged-accessible,
unprivileged-inaccessible, and non-executable regardless of the MPU
configuration.  Third, and most importantly, the MPU does not have the
execute-only permission necessary to support XOM; an MPU region is
executable only if it is configured as both readable and executable.

\subsection{Debug Support}
\label{sec:bg:dwt}

Debug support is another processor feature that ARMv7-M and ARMv8-M
devices can optionally support.  Of all components in the architecture's
debug support, we focus on the DWT
unit~\cite{ARMv7-M:Manual,ARMv8-M:Manual} which provides groups of debug
registers called \emph{DWT
comparators} that support instruction/data address matching, PC value
tracing, cycle counters, and many other functionalities.
Most importantly, a DWT comparator enables monitoring of read accesses over a
specified address range; if the processor reads from
or writes to an address within a specified range, the DWT comparator
will halt the software execution or generate a debug monitor exception.
If, instead, the access does not fall into the specified range,
execution proceeds as normal, and performance is unaffected. When
multiple DWT comparators are configured for data address range matching,
an access that hits any of them will trap.

On ARMv7-M, a DWT comparator can be configured to match an address range
by programming its base address with a mask that specifies a power-of-two
range size~\cite{ARMv7-M:Manual}.  ARMv8-M implements DWT address range
matching by using two consecutively numbered DWT
comparators~\cite{ARMv8-M:Manual},
where the first one specifies the lower bound of the address
range and the second one specifies the upper bound.

\section{Threat Model and System Assumptions}
\label{sec:threat}

We assume a buggy but unmalicious application running on an embedded
device with memory safety vulnerabilities that allow a
remote attacker to read or write arbitrary memory locations.  The
attacker wants to either steal proprietary application code
for purposes like reverse engineering or learn the application code
layout in order to launch code reuse attacks such as
Return-into-libc~\cite{Ret2Libc:RAID11}
and
Return-Oriented Programming (ROP)~\cite{ROP:TOISS12} attacks.
Physical and offline attacks are
out of scope as we believe such attacks can be stopped by orthogonal
defenses~\cite{IoT:DAC15,IoTSec:CN18}.
Our threat model also assumes the application code and data is
diversified, using techniques such as those in
EPOXY~\cite{EPOXY:Oakland17}.  Therefore, remotely tricking the buggy
application into reading its code content becomes a reasonable choice
for the attacker.

We assume that the target embedded device supports MPU
and DWT with enough configurable MPU regions and DWT comparators.  We
assume that the device is running a single bare-metal application
statically linked with libraries, boot sequences, and exception
handlers.  The application is assumed to run in privileged mode, as
Section~\ref{sec:bg:isa} dictates.  For ARMv8-M devices with TrustZone-M,
the application is assumed to reside in the non-secure world, while
software in the secure world is trusted.

\section{Design}
\label{sec:design}

\begin{figure}[tb]
  \centering
  \resizebox{0.8\columnwidth}{!}{%
    \includegraphics{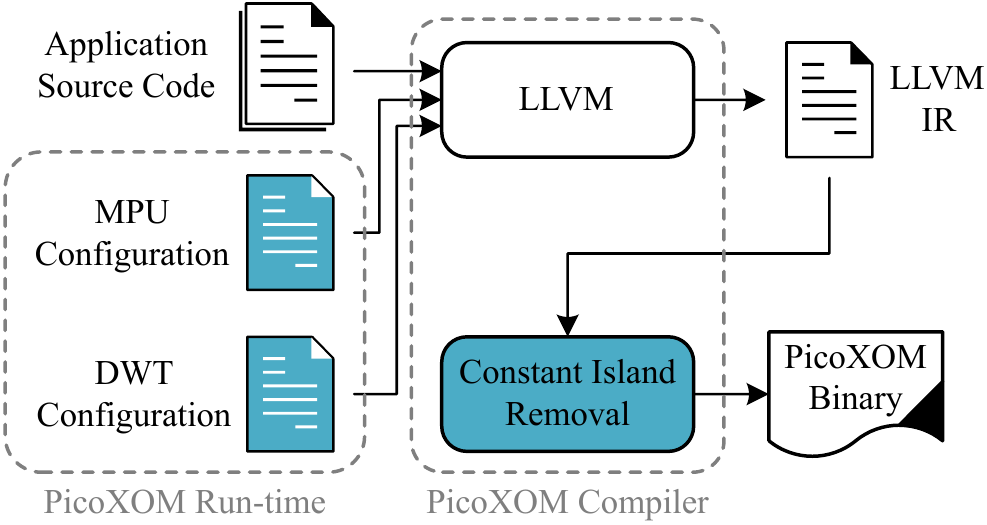}
  }
  \caption{{\System} Workflow.  {\System} components are shown in blue.}
  \label{fig:arch}
\end{figure}

Figure~\ref{fig:arch} shows {\System}'s overall design.
{\System} consists of three components that together implement a strong and
efficient XOM on ARM embedded devices.  First, {\System} uses a
specially-configured DWT configuration to detect read accesses to
program code.  Second, it utilizes a special MPU configuration that
prevents write access to the code region and prevents writeable memory
from being executable.  Third, it employs a small change to the LLVM
compiler~\cite{LLVM:CGO04} to eliminate constant data embedded within
the code region.

To use {\System}, embedded application developers merely compile
their code with the {\System} compiler and install it on their embedded
ARM device.  On boot, the {\System} run-time configures MPU regions and
DWT comparators using {\System}'s MPU and DWT configurations and
then passes control to the compiled embedded software.

\subsection{W{\XOR}X with MPU}
\label{sec:design:mpu}

{\System} requires that memory either be writeable or executable but
not both i.e., the W{\XOR}X policy~\cite{NoExec:PaX00}; otherwise,
an attacker could simply inject code or overwrite code to achieve
arbitrary code execution.
To enforce W{\XOR}X, {\System}
configures the MPU regions at device boot time so that the
code region is readable and executable, read-only data is read-only,
and RAM regions are readable and writable.  Note that the MPU
\emph{cannot} configure memory to be executable but unreadable;
the MPU can configure a memory region as executable only if it is also
configured as readable~\cite{ARMv7-M:Manual,ARMv8-M:Manual}.

{\System} runs application code in privileged mode and configures
a background MPU region to
allow read and write access to the remainder of the address space such
as peripherals.  This, however, leaves critical memory-mapped system
registers in the PPB (such as MPU configuration registers and
{\tt VTOR}) open to modifications, which can be
leveraged by an attacker to turn off MPU protections or, even worse,
implant a custom exception handler.  Section~\ref{sec:design:dwt}
discusses how {\System} prevents such cases.

\subsection{R{\XOR}X with DWT}
\label{sec:design:dwt}

{\System} leverages ARM's DWT comparators to watch over the whole
code region for read accesses.  As Section~\ref{sec:bg:dwt} states,
each (pair) of DWT comparators available on an ARM microcontroller
can be configured to generate a debug monitor exception when a
memory access of a specified type to an address within a specified
range occurs.  {\System} therefore uses one (pair) of the available DWT
comparators as follows:

\begin{enumerate}
\item
  At device boot time, {\System} configures a DWT comparator register
  (say {\tt DWT\_COMP<n>}) to hold the lower bound of the code region.
\item
  {\System} then sets the address-matching range by either writing the
  upper bound of the code region to the next DWT comparator register
  {\tt DWT\_COMP<n+1>} (for ARMv8-M) or writing the correct mask to the
  corresponding DWT mask register {\tt DWT\_MASK<n>} (for ARMv7-M).
\item
  {\System} enables the DWT comparator (pair) by configuring the DWT function
  register {\tt DWT\_FUNC<n>} for data address reads.  For ARMv8-M
  devices, {\tt DWT\_FUNC<n+1>} is also configured in order to form address
  range matching.
\item
  Finally, {\System} enables the debug monitor exception by setting the
  {\tt MON\_EN} bit (bit 16) of the Debug Exception and Monitor Control
  Register {\tt DEMCR}.
\end{enumerate}

With a DWT comparator (pair) set up for monitoring read accesses to the code
region, R{\XOR}X is effectively enforced.  However, as
Section~\ref{sec:design:mpu}
stated, the DWT registers and {\tt DEMCR} are also memory-mapped system
registers which could be modified by vulnerable application code.  An
attacker could leverage a buffer overflow vulnerability to
reconfigure the debug registers to neutralize {\System}.

We can address the issue in two ways.  One approach is to break
the assumption that {\System} runs everything in privileged mode.  As
code running in unprivileged mode has no access to the PPB region
regardless of the MPU configuration, the system registers that
{\System} must protect (e.g., MPU configuration registers, DWT
registers, {\tt DEMCR}, and {\tt VTOR}) are
all in the PPB region and therefore inherently safe from unprivileged
tampering.  However, this approach requires {\System} to implement
system calls that support privileged operations which application code
could previously perform, incurring expensive context switching
between privilege modes.  The other approach is to use extra (pairs of) DWT
comparators to prevent writes to critical system registers.  For
example, on ARMv7-M, we can configure one DWT comparator to
write-protect the
System Control Block SCB ({\tt 0xE000ED00} -- {\tt 0xE000ED8F}) and
{\tt DEMCR} ({\tt 0xE000EDFC}) by setting the lower bound and the size to
{\tt 0xE000ED00} and 256 bytes, respectively.  Since MPU
configuration registers are in the SCB, they are protected as well.  DWT
registers on ARMv7-M reside in a separate range ({\tt 0xE0001000} --
{\tt 0xE0001FFF}), so we can use another DWT comparator to
write-protect that range.

\subsection{Constant Island Removal}
\label{sec:design:xo}

\begin{figure}[tb]
  \begin{minipage}{0.44\columnwidth}
\begin{lstlisting}[language={[Arm]Assembler},framexleftmargin=-2.4em]
    ldr   r0,=L
    ...
 L: .word 0x12345678
\end{lstlisting}
  \end{minipage}
  \begin{minipage}{0.08\columnwidth}
    \hfill
    \resizebox{0.8\columnwidth}{!}{%
      \includegraphics{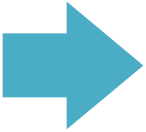}
    }
  \end{minipage}
  \begin{minipage}{0.44\columnwidth}
\begin{lstlisting}[language={[Arm]Assembler}]
movw r0,#0x5678
movt r0,#0x1234
...
\end{lstlisting}
  \end{minipage}
  \caption{Constant Island Removal of a Load Constant}
  \label{fig:ldr-pc}
\end{figure}

\begin{figure}[tb]
  \begin{minipage}{0.36\columnwidth}
\begin{lstlisting}[language={[Arm]Assembler},framexleftmargin=-2.4em]
    tbb   [pc,r2]
L0: .byte (L1-L0)/2
    .byte (L2-L0)/2
    .byte (L3-L0)/2
    ...
L1: ...
L2: ...
L3: ...
    ...
\end{lstlisting}
  \end{minipage}
  \begin{minipage}{0.08\columnwidth}
    \hfill
    \resizebox{0.8\columnwidth}{!}{%
      \includegraphics{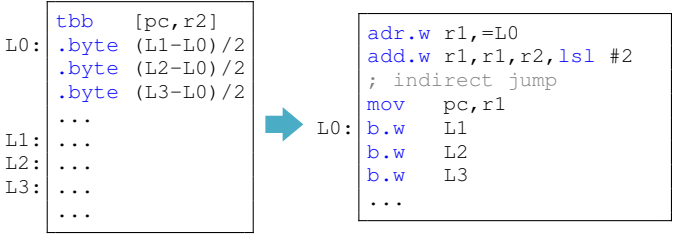}
    }
  \end{minipage}
  \begin{minipage}{0.52\columnwidth}
\begin{lstlisting}[language={[Arm]Assembler},framexleftmargin=-2.4em]
    adr.w r1,=L0
    add.w r1,r1,r2,lsl #2
    ; indirect jump
    mov   pc,r1
L0: b.w   L1
    b.w   L2
    b.w   L3
    ...
\end{lstlisting}
  \end{minipage}
  \caption{Constant Island Removal of a Jump-Table Jump}
  \label{fig:tbb}
\end{figure}

By default, ARM compilers generate code that has constant data
embedded in the code region (so-called ``constant islands'').
Since {\System} prevents the code from reading these constant islands,
these programs will fail to execute when used with {\System}.
{\System} therefore transforms these programs so that all data within
the program is stored outside of the code region.

We have identified two cases of constant islands generated by
LLVM's ARM code generator: \emph{load constants} and \emph{jump-table jumps}.
Figures~\ref{fig:ldr-pc} and~\ref{fig:tbb} show examples of the two
cases, respectively, as well as their corresponding execute-only
versions to which {\System} transforms them.  Specifically, in the left part of
Figure~\ref{fig:ldr-pc}, a load constant instruction loads a constant
from a PC-relative memory location {\tt L} into register {\tt r0}.
Such instructions are
usually generated to quickly load an irregular constant in light of the
limited immediate encoding scheme of the Thumb instruction
set~\cite{ARMv7-M:Manual,ARMv8-M:Manual}.
{\System}
transforms such load constants into {\tt MOVW} and
{\tt MOVT} instructions that encode the 32-bit constant in two 16-bit
immediates, as the right part of
Figure~\ref{fig:ldr-pc} shows.  Jump-table jump instructions ({\tt TBB} and
{\tt TBH})~\cite{ARMv7-M:Manual,ARMv8-M:Manual} are used to implement
large switch statements; the second
register operand ({\tt r2} in Figure~\ref{fig:tbb}) serves as an index
into a jump table pointed to by the first register operand ({\tt pc} in
Figure~\ref{fig:tbb}), and a byte/half-word offset is loaded from the
jump table to add to the program counter ({\tt pc}) to calculate the
target of the jump.
Optimizing compilers like GCC and LLVM usually select {\tt pc} as the
first register operand in order to reduce register pressure, forcing
the jump table
to be located next to the jump-table jump itself.  {\System}
transforms such jump-table jumps into instruction sequences like that shown
in the right part of Figure~\ref{fig:tbb}; it encodes the original
jump table's contents into a sequence of branch instructions and expands
the jump-table jump into a few explicit instructions that calculate
which branch instruction to jump to and perform an indirect jump.

\section{Implementation}
\label{sec:impl}

We built our {\System} prototype for the ARMv7-M architecture.  Our
prototype provides MPU and DWT configurations as a run-time component
written in C and executed at the end of the device boot sequence.
We implemented constant island removal as a
simple intermediate representation (IR) pass in the LLVM 10.0
compiler~\cite{LLVM:CGO04}.  The constant island removal pass
simply uses the existing {\tt -mexecute-only} option in
LLVM's Clang front-end and passes it along to the link-time optimization
(LTO) code generator.  Our prototype runs the constant island removal
pass when linking the IR of the application, libraries (e.g., newlib and
compiler-rt), and MPU and DWT configurations; this
ensures that all code has no constant islands.  Our prototype adds
88~source lines of C++ code to LLVM and has 177~source
lines of C~code in the {\System} run-time.  We leave the {\System}
implementation on ARMv8-M for future work.

Different ARM microcontrollers support different numbers of MPU
regions and DWT comparators, and the maximum ranges of their
DWT comparators may vary.
Our prototype runs on an STM32F469
Discovery board which supports up to 8~MPU
regions~\cite{STM32CortexM4:Manual} and 4~DWT
comparators~\cite{STM32F469I-DISCO:Manual}.  Each DWT
comparator can only watch over a maximum address range of 32~KB
(a maximal mask value of 15), limiting
our prototype to the following two options:

\begin{inparaenum}
\item
  Use all 4~DWT comparators to support a maximum code size of
  128~KB; the application must run in unprivileged mode in order
  for the critical system registers to be write-protected.

\item
  Configure one DWT comparator to write-protect the DWT registers
  ({\tt 0xE0001000} -- {\tt 0xE0001FFF}) and another to
  write-protect the SCB ({\tt 0xE000ED00} -- {\tt 0xE000ED8F}) and
  {\tt DEMCR} ({\tt 0xE000EDFC}). This protects a maximum code size
  of 64~KB using the remaining 2~DWT comparators.
\end{inparaenum}

To accommodate a wider range of applications on our board with less
performance loss, our prototype automatically chooses one option over
the other based on the application code size.  It rejects an application
if the code size exceeds our board's 128~KB limit.

While our {\System} prototype only supports single bare-metal
embedded applications, {\System} can also support multiple applications
running on an embedded real-time operating system (RTOS) such as Amazon
FreeRTOS~\cite{FreeRTOS:Amazon}.  On embedded systems, the application
and RTOS kernel code is linked into a single shared code
segment.  {\System} can protect this code segment with little adaptation.

\section{Evaluation}
\label{sec:eval}

We evaluated {\System} on our STM32F469 Discovery
board~\cite{STM32F469I-DISCO:Manual} which has an ARM Cortex-M4
processor implementing the ARMv7-M architecture that can run as fast as
180~MHz.  The board comes with 2~MB of
flash memory, 384~KB of SRAM, and 16~MB of SDRAM, and has an LCD screen
and a microSD card slot.  We configured the
board to run at its fastest speed to understand the maximum impact
that {\System} can incur on performance.

%
%
\begin{table}[ptb]
\caption{Performance Overhead on BEEBS}
\label{tbl:perf-beebs}
\centering
\sffamily
\footnotesize{
\resizebox{\columnwidth}{!}{
\begin{tabular}{@{}lrr|lrr@{}}
\toprule
  & {\bf Baseline} & {\bf {\System}} & & {\bf Baseline} & {\bf {\System}} \\
  & {\bf (ms)} & {\bf ($\times$)} & & {\bf (ms)} & {\bf ($\times$)} \\
\midrule
  aha-compress & 821 & 1.0000 & nettle-arcfour & 814 & 1.0000 \\
  aha-mont64 & 856 & 0.9988 & picojpeg & 43,864 & 1.0027 \\
  bubblesort & 4,392 & 1.0000 & qrduino & 40,877 & 1.0030 \\
  crc32 & 956 & 1.0000 & rijndael & 70,024 & 1.0018 \\
  ctl-string & 630 & 1.0000 & sglib-arraybin... & 808 & 1.0000 \\
  ctl-vector & 786 & 0.9987 & sglib-arrayhea... & 1,039 & 1.0000 \\
  cubic & 35,140 & 1.0005 & sglib-arrayqui... & 735 & 1.0000 \\
  dijkstra & 36,582 & 1.0000 & sglib-dllist & 1,800 & 1.0000 \\
  dtoa & 631 & 1.0127 & sglib-hashtable & 1,302 & 1.0000 \\
  edn & 3,167 & 1.0003 & sglib-listinsert... & 2,030 & 1.0000 \\
  fasta & 16,900 & 0.9999 & sglib-listsort & 1,265 & 1.0008 \\
  fir & 16,048 & 1.0000 & sglib-queue & 1,177 & 1.0000 \\
  frac & 5,858 & 1.0323 & sglib-rbtree & 4,808 & 1.0025 \\
  huffbench & 20,682 & 0.9995 & slre & 2,761 & 0.9873 \\
  levenshtein & 2,685 & 1.0000 & sqrt & 38,506 & 1.0748 \\
  matmult-float & 1,150 & 0.9991 & st & 20,906 & 1.0252 \\
  matmult-int & 4,532 & 1.0000 & stb\_perlin & 5,132 & 1.0306 \\
  mergesort & 24,353 & 1.0062 & trio-snprintf & 697 & 1.0100 \\
  nbody & 128,126 & 1.0090 & trio-sscanf & 1,064 & 0.9915 \\
  ndes & 2,039 & 0.9995 & whetstone & 112,754 & 1.0092 \\
  nettle-aes & 5,687 & 0.9998 & wikisort & 113,195 & 1.0008 \\
\midrule
  {\bf Min ($\times$)} & \multicolumn{4}{r}{} & 0.9873 \\
  {\bf Max ($\times$)} & \multicolumn{4}{r}{} & 1.0748 \\
  {\bf Geomean ($\times$)} & \multicolumn{4}{r}{} & 1.0046 \\
\bottomrule
\end{tabular}
}}
\end{table}

To evaluate {\System}'s performance and code size overhead, we used the
BEEBS~\cite{BEEBS:ArXiv13} and CoreMark-Pro~\cite{CoreMark-Pro}
benchmark suites and five embedded applications (FatFs-RAM,
FatFs-uSD, LCD-Animation, LCD-uSD, and PinLock).  {\bf BEEBS} targets
energy consumption measurement for embedded platforms and is widely used
in evaluating embedded systems including uXOM~\cite{uXOM:UsenixSec19},
the state-of-the-art XOM implementation on ARM microcontrollers.  It
consists of a wide range of programs characterizing different
workloads seen on embedded systems, including AES encryption, data
compression, and matrix multiplication.
Of all 80 benchmarks in BEEBS,
we picked 42 benchmarks that have an execution time
longer than 500 milliseconds when executed for 10,240 iterations.
{\bf CoreMark-Pro} is a
processor benchmark suite that works on both
high-performance processors and low-end microcontrollers, featuring five
integer benchmarks (e.g., JPEG image compression, XML parser, and
SHA-256) and four floating-point benchmarks (e.g., fast Fourier
transform and neural network) that stress the
CPU and memory.  {\bf FatFs-RAM} and {\bf FatFs-uSD} operate a FAT file
system on SDRAM and an SD card, respectively. {\bf LCD-Animation}
displays a single animated picture loaded from an SD card.
{\bf LCD-uSD} displays multiple static pictures from an SD card with
fading transitions.
{\bf PinLock} simulates a smart lock reading
user input from a serial port and deciding whether to unlock
(send an I/O signal) based on whether the SHA-256 hashed input matches
a precomputed hash.  The above five applications
represent real-world use cases of embedded devices and were also used
to evaluate previous work~\cite{EPOXY:Oakland17,ACES:UsenixSec18,uRAI:NDSS20}.

We used the LLVM compiler infrastructure~\cite{LLVM:CGO04} to compile
benchmarks and applications into the default non-XO format,
with MPU and DWT disabled; this is our baseline.  We then used
{\System}'s configuration, i.e. enabling MPU, DWT, and constant island
removal.  Note that with {\System}, none
of the benchmarks and applications exceeds the code size limitation
(128 KB) on our board.  Only {\tt cjpeg-rose7-preset} in CoreMark-Pro has a
code size larger than 64~KB and thereby has to run in unprivileged mode;
nevertheless, it does not require source code modifications as it does
not perform privileged operations.

\subsection{Performance}
\label{sec:eval:perf}

%
%
\begin{table}[ptb]
\caption{Performance Overhead on CoreMark-Pro}
\label{tbl:perf-coremark-pro}
\centering
\sffamily
\footnotesize{
\resizebox{\columnwidth}{!}{
\begin{tabular}{@{}lrr|lrr@{}}
\toprule
  & {\bf Baseline} & {\bf {\System}} & & {\bf Baseline} & {\bf {\System}} \\
  & {\bf (ms)} & {\bf ($\times$)} & & {\bf (ms)} & {\bf ($\times$)} \\
\midrule
  cjpeg-rose7-... & 10,200 & 1.0001 & parser-125k & 12,363 & 1.0012 \\
  core & 83,160 & 0.9918 & radix2-big-64k & 21,955 & 0.9961 \\
  linear\_alg-... & 22,962 & 1.0000 & sha-test & 25,463 & 0.9995 \\
  loops-all-... & 33,830 & 0.9995 & zip-test & 23,227 & 1.0000 \\
  nnet\_test & 282,398 & 1.0017 & & & \\
\midrule
  {\bf Min ($\times$)} & \multicolumn{4}{r}{} & 0.9918 \\
  {\bf Max ($\times$)} & \multicolumn{4}{r}{} & 1.0017 \\
  {\bf Geomean ($\times$)} & \multicolumn{4}{r}{} & 0.9989 \\
\bottomrule
\end{tabular}
}}
\end{table}

\begin{figure}[tb]
  \centering
  \resizebox{1.0\columnwidth}{!}{%
    \includegraphics{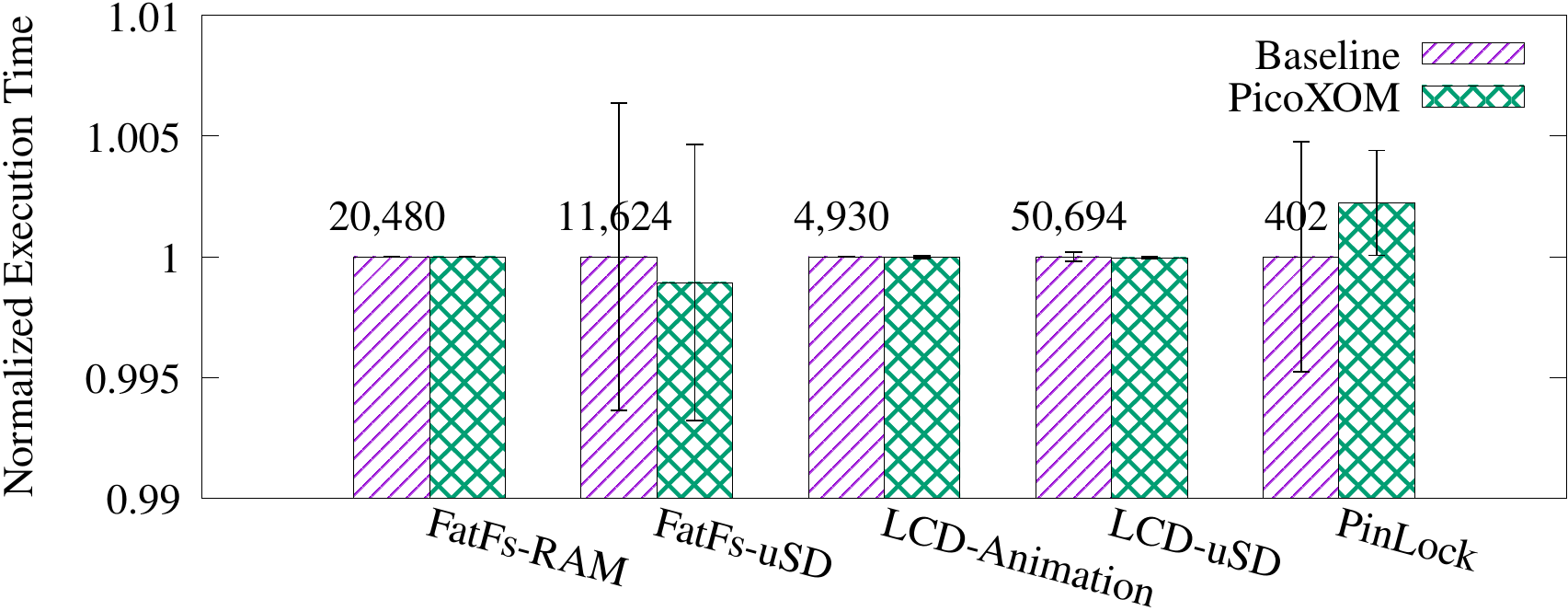}
  }
  \caption{Performance Overhead on Real-World Applications}
  \label{fig:perf-apps}
\end{figure}

We measured {\System}'s performance on our
benchmarks and applications.  We configured each BEEBS benchmark
to print the time, in milliseconds, for executing its workload 10,240
times.  We ran each BEEBS benchmark 10~times and report the average
execution time.  Each CoreMark-Pro benchmark
is pre-programmed to print out the execution time in a similar way; the
difference is that we configure each benchmark to run a minimal number
of iterations so that the program takes at least 10 seconds to run for
each experimental trial.  Again, we ran each benchmark 10~times and
report the average execution time.
For the real-world applications, we ran FatFs-RAM 10~times
and report the average execution time.  The other
applications exhibit higher variance in their execution times as
they access peripherals like an SD card, an LCD
screen, and a serial port, so we ran them
20~times and report the average with a standard deviation.
All other programs exhibit a standard deviation of zero.

Tables~\ref{tbl:perf-beebs} and~\ref{tbl:perf-coremark-pro}
and Figure~\ref{fig:perf-apps} present {\System}'s performance on BEEBS,
CoreMark-Pro, and the five real-world applications, respectively;
Figure~\ref{fig:perf-apps} shows
baseline execution time in milliseconds on top of the Baseline bars.
Overall, {\System} incurs negligible
performance overhead of 0.33\%: 0.46\% on BEEBS with a maximum of 7.48\%,
$-$0.11\% on CoreMark-Pro with a maximum of 0.17\%, and 0.02\% on
the applications with a maximum of 0.22\%.  Thirteen programs exhibit a
minor speedup with {\System}.  We re-ran our experiments with the MPU
and
DWT disabled so that the only change to performance is due to
constant island removal and the alignment of the code segment
(the DWT on ARMv7-M requires the monitored address range to be aligned
by its power-of-two size).
In this configuration, we observed the same speedups, so either
constant island removal and/or code alignment is causing the slight
performance improvement.

\subsection{Code Size}
\label{sec:eval:mem}

%
%
\begin{table}[ptb]
\caption{Code Size Overhead on BEEBS}
\label{tbl:mem-beebs}
\centering
\sffamily
\footnotesize{
\resizebox{\columnwidth}{!}{
\begin{tabular}{@{}lrr|lrr@{}}
\toprule
  & {\bf Baseline} & {\bf {\System}} & & {\bf Baseline} & {\bf {\System}} \\
  & {\bf (bytes)} & {\bf ($\times$)} & & {\bf (bytes)} & {\bf ($\times$)} \\
\midrule
  aha-compress & 30,164 & 1.0646 & nettle-arcfour & 29,988 & 1.0649 \\
  aha-mont64 & 31,236 & 1.0624 & picojpeg & 36,620 & 1.0599 \\
  bubblesort & 29,868 & 1.0650 & qrduino & 37,228 & 1.0529 \\
  crc32 & 29,804 & 1.0654 & rijndael & 37,460 & 1.0516 \\
  ctl-string & 30,668 & 1.0631 & sglib-arraybin... & 29,828 & 1.0654 \\
  ctl-vector & 30,892 & 1.0624 & sglib-arrayhea... & 29,956 & 1.0651 \\
  cubic & 42,428 & 1.0329 & sglib-arrayqui... & 30,036 & 1.0649 \\
  dijkstra & 30,220 & 1.0644 & sglib-dllist & 30,364 & 1.0641 \\
  dtoa & 36,204 & 1.0552 & sglib-hashtable & 30,164 & 1.0644 \\
  edn & 30,940 & 1.0633 & sglib-listinsert... & 30,052 & 1.0649 \\
  fasta & 29,956 & 1.0650 & sglib-listsort & 30,100 & 1.0648 \\
  fir & 29,884 & 1.0651 & sglib-queue & 29,988 & 1.0650 \\
  frac & 30,468 & 1.0626 & sglib-rbtree & 30,564 & 1.0639 \\
  huffbench & 30,988 & 1.0628 & slre & 32,284 & 1.0603 \\
  levenshtein & 30,140 & 1.0647 & sqrt & 30,372 & 1.0641 \\
  matmult-float & 30,108 & 1.0644 & st & 31,124 & 1.0602 \\
  matmult-int & 30,060 & 1.0650 & stb\_perlin & 31,140 & 1.0627 \\
  mergesort & 30,852 & 1.0604 & trio-snprintf & 33,724 & 1.0675 \\
  nbody & 30,684 & 1.0633 & trio-sscanf & 34,156 & 1.0668 \\
  ndes & 31,028 & 1.0630 & whetstone & 40,164 & 1.0371 \\
  nettle-aes & 31,756 & 1.0614 & wikisort & 34,332 & 1.0541 \\
\midrule
  {\bf Min ($\times$)} & \multicolumn{4}{r}{} & 1.0329 \\
  {\bf Max ($\times$)} & \multicolumn{4}{r}{} & 1.0675 \\
  {\bf Geomean ($\times$)} & \multicolumn{4}{r}{} & 1.0614 \\
\bottomrule
\end{tabular}
}}
\end{table}

\begin{figure}[tb]
  \centering
  \resizebox{1.0\columnwidth}{!}{%
    \includegraphics{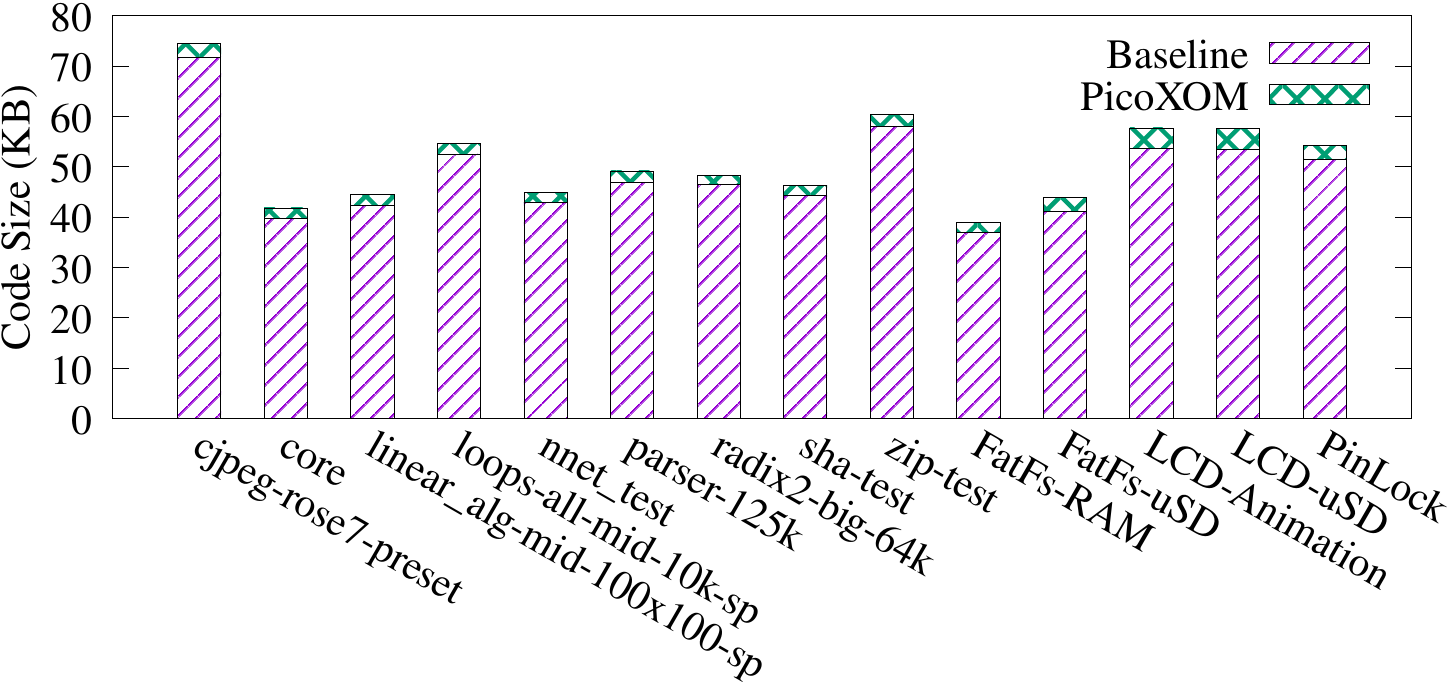}
  }
  \caption{Code Size Overhead on CoreMark-Pro and Real-World Applications}
  \label{fig:mem-coremark-pro-apps}
\end{figure}

We measured the code size of benchmarks and applications by using the
{\tt size} utility on generated binaries and collecting the {\tt .text}
segment size.

Table~\ref{tbl:mem-beebs} and Figure~\ref{fig:mem-coremark-pro-apps}
show the baseline code size and the overhead
incurred by {\System} on BEEBS, CoreMark-Pro, and the five real-world
applications, respectively.  On average, {\System} increases the code
size by 6.14\% on BEEBS, 4.39\% on CoreMark-Pro, and 6.52\% on the
real-world applications, with a 5.89\% overall overhead.
We studied {\System}'s code size overhead and
discovered that constant island removal caused
the majority of the code size overhead, especially for programs with
relatively large code bases like
CoreMark-Pro.  In fact, the additional code that sets up the MPU
and DWT only contributes a minor part of the overhead (1.22\% and 0.53\%
on average, respectively).

\section{Related Work}
\label{sec:related}

%
%

Two other XOM implementations exist for ARM microcontrollers.
uXOM~\cite{uXOM:UsenixSec19}
provides XOM for ARM Cortex-M systems by transforming loads into
special unprivileged load
instructions and configuring the MPU to make the code region unreadable by
unprivileged loads.  uXOM similarly transforms stores to
protect the memory-mapped MPU configuration registers.  Since
some loads and stores do not have unprivileged counterparts,
transforming them requires the compiler to insert additional instructions,
causing the majority of uXOM's overhead.  {\System} is more
efficient in both performance (0.33\% compared to uXOM's 7.3\%) and code
size (5.89\% compared to uXOM's 15.7\%) as no such transformation is needed.
A trade-off for {\System} is the code size limit on some ARMv7-M
devices; we envision no such limit on ARMv8-M.
PCROP~\cite{STM32F4:PCROP:Manual} is a programmable feature of the flash
memory which prevents the flash memory from being read out and modified by
application code but still allows code in the flash memory to execute.
However, PCROP is only available on some
STMicroelectronics devices and cannot be used for other types of memory.
In contrast, {\System} relies on the MPU
and DWT features~\cite{ARMv7-M:Manual,ARMv8-M:Manual} which can
be found on most conforming devices
and can protect code stored in any type of memory.

%
%

Hardware-assisted XOM has been explored on other architectures.
The AArch64~\cite{ARMv8-A:Manual} and RISC-V~\cite{RISC-V:Priv:Manual}
page tables natively support XO permissions.
NORAX~\cite{NORAX:Oakland17} enables XOM for
commercial-off-the-shelf binaries on AArch64 that have constant islands
using static binary
instrumentation and runtime monitoring.
Various approaches~\cite{HideM:CODASPY15,Readactor:Oakland15,%
ExOShim:ICCWS16,KHide:CNS16,XOM-Switch:BlackHatAsia18,IskiOS:ArXiv19} leverage
features of the MMU on Intel x86 processors~\cite{X86:Intel:Manual}
to implement XOM.
None of these approaches are applicable on ARM embedded devices
lacking an MMU.  Lie et al.~\cite{XOM:ASPLOS00} proposed an
architecture with memory encryption to mimic XOM, but it only
provides probabilistic guarantees and cannot be directly applied to
current embedded systems.
Compared to solutions for systems lacking native hardware XOM support,
{\System} is faster as it has nearly no overhead.

%
%

Software can emulate XOM.  XnR~\cite{XnR:CCS14}
maintains a sliding window of currently executing code pages and
keeps only these pages accessible.
It still allows read accesses to a subset of code pages
and may incur higher overhead for a smaller sliding
window size due to frequent page permission changes.
LR$^2$~\cite{LR2:NDSS16} and kR\^{}X~\cite{kR^X:EuroSys17} instrument
all load instructions to prevent them from reading the code
segment.  While these software XOM approaches can generally be ported
to embedded devices,
they can be bypassed by attacker-manipulated control flow and
are less efficient than hardware-assisted XOM~\cite{uXOM:UsenixSec19}.

%
%

There are also methods of hardening embedded systems.
Early versions of SAFECode~\cite{DKAL:TECS05} enforced spatial and
temporal memory safety on embedded applications, and
nesCheck~\cite{nesCheck:ASIACCS17} uses static analysis to build
spatial memory safety for simple nesC~\cite{nesC:PLDI03} applications
running on TinyOS~\cite{TinyOS:ASPLOS00}.
{\System} enforces weaker protection than
memory safety but supports arbitrary C programs (unlike SAFECode and nesCheck)
and does not rely on heavy static analysis like nesCheck.
RECFISH~\cite{RECFISH:ECRTS19},
$\mu$RAI~\cite{uRAI:NDSS20}, and Silhouette~\cite{Silhouette:UsenixSec20}
mitigate control-flow hijacking attacks on embedded systems.  They
protect forward-edge control flow using coarse-grained CFI~\cite{CFI:CCS05} and
backward-edge control flow by using either a protected shadow
stack~\cite{SoK:SS:Oakland19} or a return address encoding mechanism.
EPOXY~\cite{EPOXY:Oakland17} randomizes the order of functions and the
location of a modified safe stack from CPI~\cite{CPI:OSDI14} to resist
control-flow hijacking attacks on bare-metal microcontrollers.  These
systems do not enforce XOM and are still vulnerable to forward-edge
corruptions; they can incorporate {\System}'s techniques to mitigate
forward-edge attacks with negligible additional overhead.

\section{Conclusions and Future Work}
\label{sec:conc}

This paper presented {\System}: a fast and novel XOM system for
ARMv7-M and ARMv8-M devices which leverages ARM's MPU and
DWT unit.
{\System} incurs an average performance overhead of 0.33\% and an average
code size
overhead of 5.89\% on the BEEBS and CoreMark-Pro benchmark suites and
five real-world applications.

In future work, we will investigate techniques to ensure that
randomization techniques utilizing {\System} are effective
against brute-force attacks.  Embedded systems have limited code placement
options for code layout randomization, motivating us to
investigate whether the entropy is sufficient and develop
techniques to strengthen code randomization if necessary.
We will also explore how to leverage debug support like DWT to enforce
other security policies with low overhead.

\section*{Acknowledgements}
\label{sec:ack}

We thank the anonymous reviewers for their insightful comments.  This work was funded
by ONR Award N00014-17-1-2996.

\bibliographystyle{IEEEtranS}
\bibliography{picoxom}

\end{document}